\documentclass[aps,prl,twocolumn,superscriptaddress]{revtex4-1}

\usepackage{graphicx}
\usepackage{epstopdf}
\usepackage{amssymb}
\usepackage{array}
\usepackage{supertabular}
\usepackage{amsmath}
\usepackage{bm}
\usepackage[hidelinks]{hyperref}

\begin{document}
	\title{Polar Kerr Effect from Time Reversal Symmetry Breaking in the Heavy Fermion Superconductor PrOs$_4$Sb$_{12}$}
	
	\author{E. M. Levenson-Falk}
	\email[Corresponding author: ]{elevenso@usc.edu}
	\affiliation{Department of Applied Physics, Stanford University, Stanford, CA 94305}
	\affiliation{Geballe Laboratory for Advanced Materials, Stanford University, Stanford, CA 94305}
	\affiliation{Department of Physics and Astronomy, University of Southern California, Los Angeles, CA 90089}
	\author{E. R. Schemm}
	\affiliation{Geballe Laboratory for Advanced Materials, Stanford University, Stanford, CA 94305}
	\author{Y. Aoki}
	\affiliation{Department of Physics, Tokyo Metropolitan University, Tokyo 192-0397}
	\author{M. B. Maple}
	\affiliation{Department of Physics and Center for Advanced Nanoscience, University of California San Diego, La Jolla, CA 92093}
	\author{A. Kapitulnik}
	\affiliation{Department of Applied Physics, Stanford University, Stanford, CA 94305}
	\affiliation{Department of Physics, Stanford University, Stanford, CA 94305}
	\affiliation{Stanford Institute for Materials and Energy Sciences, SLAC National Accelerator Laboratory, 2575 Sand Hill Road, Menlo Park, California 94025, USA}

	\begin{abstract}
		We present polar Kerr effect measurements of the filled skutterudite superconductor PrOs$_4$Sb$_{12}$.  Simultaneous AC susceptibility measurements allow us to observe the superconducting transition under the influence of heating from the optical beam.  A nonzero Kerr angle $\theta_K$ develops below the superconducting transition, saturating at $\sim 300$ nrad at low temperatures.  This result is repeated across several measurements of multiple samples. By extrapolating the measured $\theta_K(T)$ to zero optical power, we are able to show that the Kerr angle onset temperature in one set of measurements is consistent with the transition to the B phase at $T_{C2}$. We discuss the possible explanations for this result and its impact on the understanding of multi-phase and inhomogeneous superconductivity in PrOs$_4$Sb$_{12}$.
	\end{abstract}
	
	\maketitle
	
Chiral superconducting phases exhibited by certain heavy fermion (HF) materials have recently attracted heightened interest as hosts for Majorana particles and other topologically ordered states \cite{Chakravarty2015,Kallin2016,Sato2016}.  Of these, the filled skutterudite PrOs$_4$Sb$_{12}$ has been proposed as a leading candidate for hosting three-dimensional Majorana fermions \cite{Kozii2016}.  PrOs$_4$Sb$_{12}$, currently the only known Pr-based HF-superconductor \cite{Bauer2002}, exhibits many interesting phenomena \cite{Maple2006}, including a field-induced ordered state \cite{Aoki2002} and two zero-field superconducting transitions, corresponding to an A phase with $T_{C1} \approx 1.85$ K, and a B phase with $T_{C2} \approx 1.72$ K \cite{Izawa2003,Vollmer2003}. Past experiments have suggested that quadrupolar order and fluctuations may be the basis of all these phases \cite{Bauer2002,Maple2006,Aoki2007}.  Theoretical models posit that such a quadrupolar superconducting state breaks time-reversal symmetry (TRS) and that the double transition arises from spin-orbit coupling \cite{Miyake2003}.  A muon spin relaxation study of PrOs$_4$Sb$_{12}$ showed evidence of TRS breaking (TRSB) appearing around the superconducting transition, thus suggesting that the superconductivity is chiral, but could not resolve whether TRSB was associated with the A phase or B phase \cite{Aoki2003}.  Whether these transitions are even distinct, and what are the possible symmetries allowed for this compound, are subjects of considerable debate \cite{Maple2006,Méasson2006,Seyfarth2006,Aoki2007}.  There is strong evidence of inhomogeneity in the superconducting state; furthermore, reducing the size of a crystal below $\sim 100 \mu$m in any dimension eliminates the transition at $T_{C1}$ \cite{Measson2008,Andraka2012}. This raises the possibility that the ``two phases" are just ordinary transitions of two different materials in the same sample. However, this inhomogeneity may simply indicate that the A phase is delicate, and requires a large crystal to exist\textemdash in this case, an inhomogeneous crystal would be composed of regions which support both A and B phases, combined with regions supporting only the B phase. This argument is bolstered by evidence that powdering the material (i.~e.~ introducing defects and impurities) suppresses the upper transition \cite{McBriarty2009}. In order to clear up this controversy, it is thus crucial to determine the exact TRS of the superconducting state below $T_{C2}$ and between $T_{C1}$ and $T_{C2}$.

In this Letter we report polar Kerr effect measurements of several PrOs$_4$Sb$_{12}$ crystals along the [001] direction using a zero-area loop Sagnac interferometer (ZALSI).  In-situ AC susceptibility measurements allow us to track the two superconducting transitions with the optical beam incident and thus accurately account for optical heating.  We find a finite Kerr angle, saturating at $\sim \pm300$ nrad at low temperature, which develops below the superconducting transition temperature after the sample has been cooled in a small symmetry-breaking magnetic field.  When cooled in zero field, the sample develops a Kerr angle with random sign and magnitude (also saturating at $\sim \pm300$ nrad), indicating the formation of TRSB domains with random direction. By measuring (on one sample) the power dependence of the temperature at which the Kerr angle onsets, we are able to extrapolate to the true onset temperature of TRSB in the limit of no sample heating. We find the onset temperature is consistent only with $T_{C2}$.  We measure several other spots on multiple samples, finding results which are consistent with this power dependence. We take this to indicate that PrOs$_4$Sb$_{12}$ has a TRSB superconducting state below $T_{C2}$, which may be the B phase of a multi-phase superconductor.

In general, a TRSB order parameter with particle-hole asymmetry will lead to complex indices of refraction for right-circular ($n_\mathrm{R}$) and left-circular ($n_{\mathrm{L}}$) polarizations that are unequal and depend on the direction of propagation of light \cite{Sauls1994}. This generates a small but finite polar Kerr effect (PKE), wherein circularly polarized light reflected from a TRSB material is phase-shifted by the Kerr angle $\pm\theta_K$, with the sign of $\theta_K$ depending on the direction of polarization.
For a multiband, TRSB superconductor with $T_c\sim 1$K measured with $\lambda \sim 1$ $\mu$m light, we expect $\theta_K \approx 0.1 - 1$ $\mu$rad \cite{Orderofmagnitude}.  Such a small PKE may be resolved using a ZALSI as described previously \cite{Xia2006,Kapitulnik2009}.  Our ZALSI apparatus, operating at $\lambda = 1.55$ $\mu$m, yields a finite $\theta_K$ only if TRS is broken, while rejecting any reciprocal effects that may happen to rotate polarization.  Since if reciprocity holds the PKE is identically zero \cite{Kapitulnik2015}, a finite PKE is an unambiguous determination of TRSB in any material system, including unconventional superconductors. Indeed, we have used ZALSI in the past to detect TRSB in Sr$_2$RuO$_4$ and in the HF-superconductors URu$_2$Si$_2$ \cite{Schemm2015} and UPt$_3$ \cite{Schemm2014}. Importantly, the ZALSI has also ruled out TRSB in the HF-superconductor CeCoIn$_5$ \cite{Schemm2017}, in mirrors \cite{Xia2006a}, and in high-$T_C$ cuprates \cite{Kapitulnik2009a}. 
	
\begin{figure}
	\includegraphics{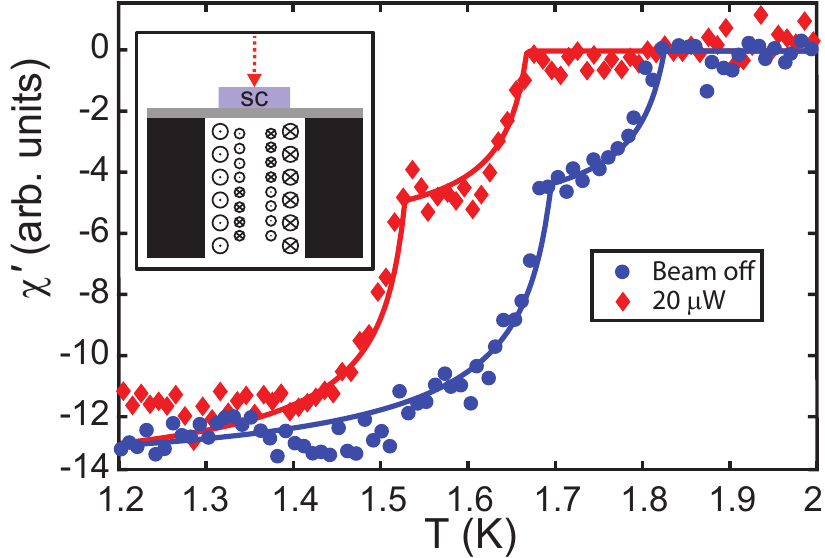}
	\caption{\label{fig:mutualinductance} (color online) AC susceptibility as a function of temperature with the optical beam off (blue circles) and at 20 $\mu$W incident power (red diamonds), measured on sample A.  Two transitions are apparent in each curve.  The solid lines are phenomenological fits emphasizing the multiple transitions.  Increasing optical power heats the sample, causing the transitions to occur at lower measured fridge temperature.  Inset: schematic of our mutual inductance apparatus (not to scale). An astatically-wound pickup coil sits inside a drive coil, with the end butted up against a thin sapphire wafer that supports the sample (labeled SC).  The optical beam (dashed red arrow) is incident from above.}
\end{figure}

We measured 5 samples in total, labeled samples A through E. These ranged in size from approximately $2\times5$ mm$^2\times 2$ mm thick to approximately $1\times1$ mm$^2\times 0.3$ mm thick. The single crystals of PrOs$_4$Sb$_{12}$ were grown in an Sb flux as described in Ref.~\cite{Bauer2001}; all samples except B were from the same growth batch, grown under identical conditions, and were measured on as-grown (100) surfaces.  X-ray diffraction measurements \cite{Bauer2002} on PrOs$_4$Sb$_{12}$ reveal that it crystallizes in the LaFe$_4$P$_{12}$-type BCC structure with a lattice parameter $a = 9.3017$ \AA \cite{Braun1980}.  These crystals typically have a residual resistivity ratio RRR $\equiv \rho($300 K$) / \rho($2 K$) \approx 30 - 50$ \cite{Bauer2002,Maple2006,Frederick2003}.  A sample is attached to a thin sapphire wafer on top of a copper stage that is mounted on the base plate of our $^3$He refrigerator.  The optical beam from the Sagnac interferometer comes down from above and reflects off the top of the sample. The mutual inductance (MI) apparatus for measuring AC susceptibility consists of an astatically-wound pickup coil inside a drive coil just underneath the sample, as shown in the inset of Fig.~\ref{fig:mutualinductance}.  We drive the outer coil at $10.054$ kHz and measure the induced voltage in the pickup.  The pickup coil is nominally balanced to provide zero signal when the sample susceptibility $\chi$ is zero and finite signal when $\chi$ is finite. In practice there is always some background signal; we measure this offset well above $T_{C1}$ and subtract it from our data.  We measure with the optical beam on, thus determining the effect of sample heating from our optical measurement.
	
PKE measurements of PrOs$_4$Sb$_{12}$ proved challenging because of the low reflectivity at the measurement wavelength ($\lambda = 1.55 \mu$m). Signal to noise ratio (SNR) could be increased by increasing optical power, which in turn resulted in heating of the sample.  To get reliable data in the limit of low optical power we measured PKE at different powers and the compared the data to AC susceptibility data taken at the same optical powers.  AC susceptibility measurements of sample A under illumination at different optical powers are shown in Fig.~\ref{fig:mutualinductance}.  With the optical beam off, we observe two transitions in the real component of the susceptibility $\chi'$ beginning at 1.83 K and 1.7 K.  These temperatures are consistent with $T_{C1}$ and $T_{C2}$ from previously reported measurements of PrOs$_4$Sb$_{12}$ \cite{Maple2006,Aoki2007}.  With the optical beam on, both transitions shift to lower temperature (as measured by a RuOx thermometer mounted to the sample stage away from the beam).  This indicates that the sample is heated by the beam and thus sits at higher temperature than the stage.  The double-transition shape of the curve remains unchanged, although at high powers a hint of the transition at 1.83 K may be seen, presumably due to uneven thermalization of the sample due to the focused heating spot.  It is the temperature of this spot that is relevant for our analysis, as the PKE measurement samples over the volume that the beam interacts with.
	
	To fit the MI data, we use the canonical susceptibility of a cylinder in a parallel field, 
	\begin{equation*}
	\chi' \sim -\left[1 - \frac{2 \lambda}{R} \frac{I_1(R / \lambda)}{I_0(R / \lambda)} \right]
	\end{equation*}
	where $R$ is the cylinder radius, $\lambda$ is the penetration depth, and $I_i$ are the modified Bessel functions \cite{Brandt1998}.  We sum the susceptibilities due to two superconducting states, $\chi = \sum_i \chi(\lambda_i)$, using a phenomenological temperature dependence for penetration depths
	\begin{equation*}
	\lambda_i = (1 - (T / T_{Ci})^4)^{-1/2}
	\end{equation*}
	The exact shape of these fits seems unimportant, as the $T_C$ values extracted do not depend strongly on the choice of functional form; however, a two-transition fit is necessary to capture the shape of the data.  The $T_C$ values extracted from fitting agree with estimates ``by eye" of the beginnings of the transitions.  We may thus use these temperatures to determine what state the sample is in at a given temperature for the Kerr measurements. We note that this functional form, which sums susceptibilities, seems to imply inhomogeneity as the source of the two transitions. However, similar curve shapes and values of $T_C$ are generated by models that assume two phases with different $\lambda_i$'s \cite{Khasanov2007}.
	
	PKE measurements were performed with no MI drive to prevent any spurious effects from the AC magnetic field.  The measurement procedure is thus: the sample is warmed far above the $T_C$, typically to 4 K.  After a short wait for thermalization, a small DC magnetic field may be applied, or the field may be left at zero.  The sample is then cooled to base temperature ($\sim 300$ mK) and the field is set to 0 (remnant field in the absence of applied field is $<\sim 3$ mG \cite{Schemm2015}, well below $H_{C1} \sim 45$ Oe \cite{Cichorek2005}).  Finally, the sample is warmed slowly and PKE measurements are performed during warm-up.
	
	\begin{figure}
		\includegraphics{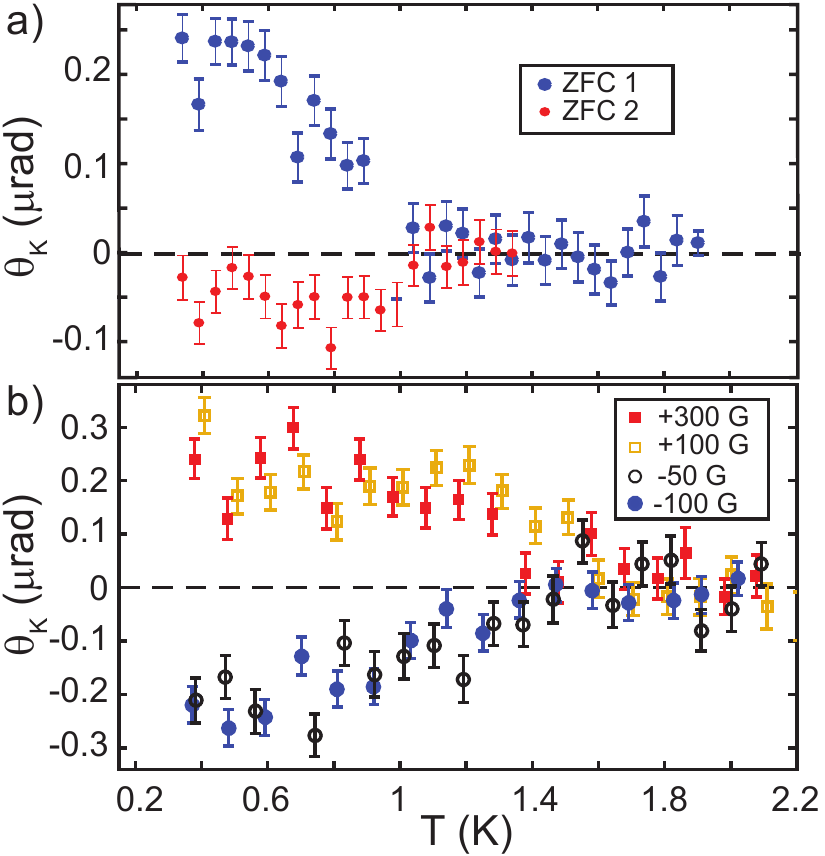}
		\caption{\label{fig:kerrfield} (color online) (a) Kerr angle as a function of temperature at 20 $\mu$W incident power after cooling in zero field (data from sample E).  Error bars represent statistical error of hundreds of data points averaged together \cite{supplement}.  Two different runs are shown; one shows a positive Kerr signal, saturating at $\sim 250$ nrad, while the other shows a small negative signal. This sample showed Kerr signals with random amplitude and sign after cooling in zero field. (b) Kerr angle at $10 \mu$W power after cooling in various fields (data from sample C).  A nonzero $\theta_K$ develops symmetrically below $\sim 1.5$ K, saturating at  $\sim 300$ nrad.  The saturating amplitude is independent of field strength.}
	\end{figure}
	
	PKE data taken on sample E after two different zero field cooldowns at $20 \mu$W incident power are shown in Fig.~\ref{fig:kerrfield}(a).  One run shows a positive signal saturating around 250 nrad, while the other shows a negative signal of roughly 50 nrad. Several ZFC runs taken on multiple samples showed Kerr angles with random sign and amplitude. This behavior, together with the field-cooled measurements discussed below, suggest that we observe the effect of finite domain structure. Similar to a finite-size ferromagnet, the TRSB sample may break into domains that are smaller than the gaussian waist of our optical beam. In this case the average signal is expected to be zero, with a standard deviation reflecting the ratio of domain width to beam waist \cite{Xia2006a}.  Occasionally a large domain may form under the beam, leading to a finite signal, but this measurement would not be repeatable. We observe exactly this behavior.
	
	Again as in a ferromagnet, the sample can be trained to form a single domain by cooling through the TRSB transition in a symmetry-breaking magnetic field aligned with the optical beam \cite{Xia2006a}.  It is important to note, however, that the field is removed at low temperature, and all data are taken while warming up in zero field.  Data taken at $10$ $\mu$W optical power after cooling sample C in field are shown in Fig.~\ref{fig:kerrfield}(b).  A nonzero $\theta_K$ develops below $\sim 1.4$ K at all fields, and saturates at $\sim \pm 300$ nrad at low temperature.  The sign of $\theta_K$ reverses with the field direction, as would be expected for trained TRSB.  The magnitude of $\theta_K$ shows no field dependence, indicating that vortex cores are likely not the source of the signal and that a 50 G field fully trains the sample.
	
	\begin{figure}
		\includegraphics{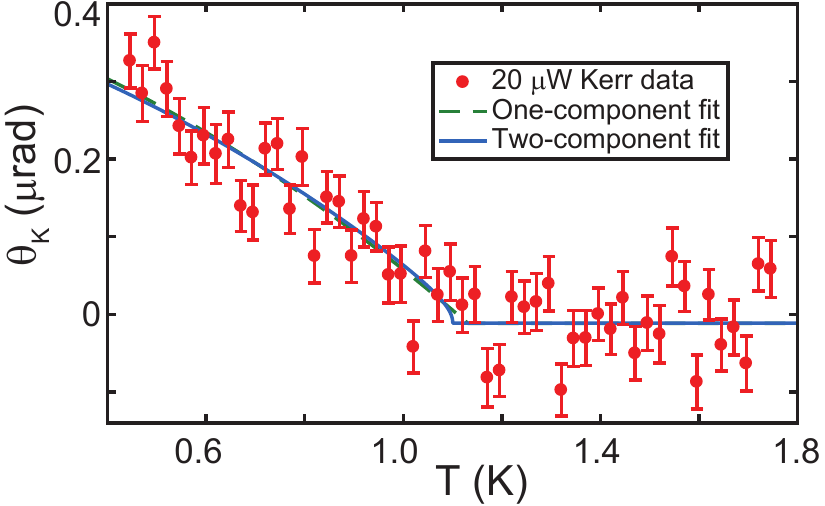}
		\caption{\label{fig:kerrfit} (color online) Averaged Kerr data taken on sample E at $20 \mu$W incident power.  Error bars are statistical error (one-$\sigma$).  Fits with two critical temperatures (solid blue line) or one (dashed green line) are overlaid.}
	\end{figure}
	
	In order to accurately extract the temperature $T_K$ at which $\theta_K$ becomes nonzero (i.~e.~ at which TRSB begins) we fit the data. An example, using $20 \mu$W data on sample E, is shown in Fig.~\ref{fig:kerrfit}. We use two phenomenological models for these fits. The first (dashed green line) assumes a single component to the order parameter: 
	\begin{equation*}
	\theta_K (T) \sim (1 - ((T+\Delta T) / T_C)^2)
	\end{equation*}
	Here, $\Delta T$ is the optical spot heating, so the measured $T_K$ is given by $T_K = T_C - \Delta T$. Either $T_{C1}$ or $T_{C2}$ may be used; the $T_K$ values extracted are nearly identical. The second model (solid blue line) assumes two components to the order parameter, i.~e.~ two critical temperatures: 
	\begin{equation*}
	\theta_K (T) \sim \sqrt{[1 - ((T+\Delta T) / T_{C1})^2][1 - ((T+\Delta T) / T_{C2})^2]}
	\end{equation*}
	Both equations fit the data well, although the two-component fit gives a smaller confidence interval for $\Delta T$ and usually has larger R-square. The one-component fit gives $T_K = 1.167 \pm 0.057$ K, while the two-component fit gives $T_K = 1.125 \pm 0.046$ K (error bars represent 95\% confidence intervals). These values are lower than the temperatures of both MI transitions (measured to be 1.672 and 1.528 K at this power); this is to be expected in the presence of optical heating, as the Kerr measurement samples only the volume heated by the optical beam while the MI measurement measures the entire sample.  A finite-element model of optical heating gives a ``hot spot" which is $200 - 750$ mK hotter than the bulk crystal, consistent with our results.
	
\begin{figure}
		\includegraphics{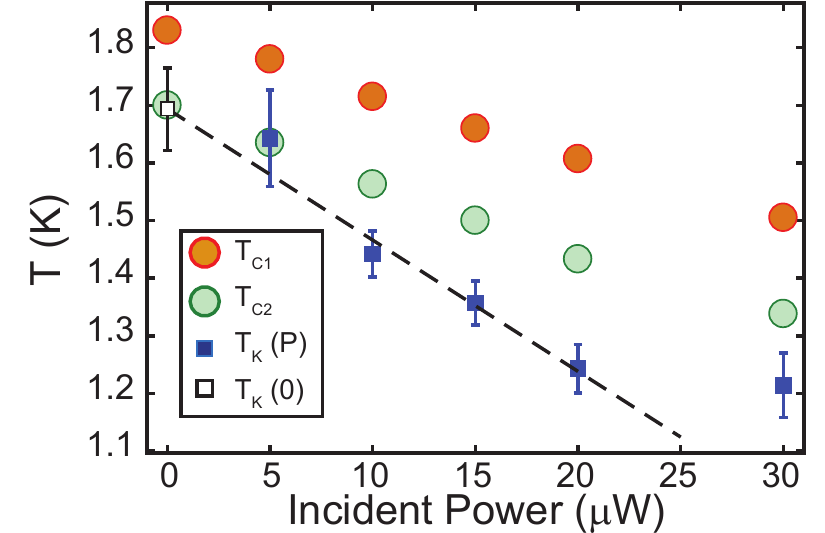}
		\caption{\label{fig:tctk} (color online) Measured values of $T_{C1}$ (orange circles), $T_{C2}$ (light green circles), and $T_K$ (blue squares) as a function of optical power on sample C.  Error bars indicate 95\% confidence intervals on the values of $T_K$ extracted from Kerr data fits.  The dashed line is a linear fit to $T_K(P)$ from $5-20$ $\mu$W, extrapolating back to $T_K(0) = 1.694 \pm 0.072$ K (white square), which is consistent with $T_{C2}$.}
	\end{figure}

	In order to determine the true value of $T_K$ (in the absence of heating) we measure Kerr data at different incident optical powers ($P$) and extract $T_K(P)$ from two-component fits as above. At low powers the SNR of our measurement is too low to consistently fit for $T_K$. However, since field-trained data is repeatable with the same amplitude and onset temperature, we may fit several data sets simultaneously to improve SNR. We then fit $T_K(P)$, weighted by the confidence interval of each extracted value, to find $T_K(P=0)$. Fit procedure details may be found in the Supplementary Material \cite{supplement}.  Results for sample C are shown in Fig.~\ref{fig:tctk}, along with measured MI transition temperatures at each power.  We fit a linear dependence to $T_K(P)$ from $P = 5$ to $20$ $\mu$W; at the highest power, $30$ $\mu$W, we expect a deviation from linear dependence as the thermal conductivities of the sample and stage have changed significantly, and the heating is sufficiently large that the Kerr data is not well-fit by the same functions.  The linear fit gives $T_K(P=0) = 1.694 \pm 0.072$ K.  These error bars represent 95\% confidence bounds \cite{supplement}.  Thus, we see that $\theta_K$ becomes nonzero at a temperature consistent with $T_{C2} = 1.7$ K and inconsistent (at the 3.8-$\sigma$ level) with $T_{C1} = 1.83$ K.  Repeating this procedure with one-component fits gives $T_K(0) = 1.74 \pm 0.097$ K, which again is more consistent with $T_{C2}$, although $T_{C1}$ is just barely within the 2-$\sigma$ window.
	
	In order to test whether $T_K = T_{C2}$ at all points\textemdash i.~e.~ to test whether inhomogeneities lead to any regions where TRSB begins at $T_{C1}$\textemdash we measured 8 different spots across 3 samples (A, C, and E) \cite{sampleBD}. Because measuring $T_K(P)$ dependence is extremely time-consuming, we instead measured at $20 \mu$W and extracted the difference between the measured $T_K$ and $T_{C2}$ at that power. This difference should depend only on the thermal conductivity of the sample and the size of the optical spot (assumed to be very similar for all samples); therefore $T_{C2}(P) - T_K(P)$ should be constant for all measurements if $T_K(0)$ is the same. We find that this value agrees in 6 of the measurements \cite{supplement}, indicating that $T_K(0) = T_{C2}$; the other 2 measurements were inconclusive due to spurious signals that corrupted the data and made fitting unreliable. We take this as moderate evidence that the superconducting states entered at $T_{C1}$ and $T_{C2}$ have different TRS, likely indicating multiphase superconductivity. 

	However, we have not fully ruled out inhomogeneities as the source of the multiple transitions. It is possible that the $T_{C1}$ transition only occurs inside a crystal, and that the surface only becomes superconducting at $T_{C2}$--indeed, the fact that reducing crystal size eliminates the upper transition would point to this possibility. Our measurement only probes down to the optical penetration depth ($\sim 100$ nm), and so is unable to determine if this is the case. In order to fully correlate the TRS and $T_C$ of the superconducting states, a measurement of $T_C$ at the optical spot is needed. This could be accomplished using a scanning probe \cite{deLozanneSPM}, a SQUID susceptometer \cite{MolerSQUID}, or by measuring optical properties such as thermoreflectance. Future experiments should focus on correlating $T_K$ and $T_C$ across many regions of a sample.

	In conclusion, we have found a finite polar Kerr effect below $T_{C2}$ in PrOs$_4$Sb$_{12}$.  The random PKE after cooling in zero field, together with a finite PKE after cooling in a small magnetic field, suggest that time-reversal symmetry is broken in PrOs$_4$Sb$_{12}$, with typical domain size that is smaller than our $\sim 10$ $\mu$m optical spot.   This finding puts strong constraints on the possible symmetries allowed to describe the superconducting state in this material system. Further study is necessary to investigate the role of inhomogeneities and determine whether the two superconducting transitions are truly distinct.
	
	\begin{acknowledgements}
	We thank Daniel Agterberg and Liang Fu for useful discussions.  Single crystal growth and characterization at UCSD was supported by the US Department of Energy, Office of Basic Energy Sciences, Division of Materials Sciences and Engineering, under Grant No. DEFG02-04-ER46105.  Kerr effect measurements at Stanford were supported under Department of Energy Contract No. DE-AC02-76SF00515. 
	\end{acknowledgements}

\bibliographystyle{apsrev4-1}
\bibliography{TRSBbib}	

\begin{thebibliography}{39}%
\makeatletter
\providecommand \@ifxundefined [1]{%
 \@ifx{#1\undefined}
}%
\providecommand \@ifnum [1]{%
 \ifnum #1\expandafter \@firstoftwo
 \else \expandafter \@secondoftwo
 \fi
}%
\providecommand \@ifx [1]{%
 \ifx #1\expandafter \@firstoftwo
 \else \expandafter \@secondoftwo
 \fi
}%
\providecommand \natexlab [1]{#1}%
\providecommand \enquote  [1]{``#1''}%
\providecommand \bibnamefont  [1]{#1}%
\providecommand \bibfnamefont [1]{#1}%
\providecommand \citenamefont [1]{#1}%
\providecommand \href@noop [0]{\@secondoftwo}%
\providecommand \href [0]{\begingroup \@sanitize@url \@href}%
\providecommand \@href[1]{\@@startlink{#1}\@@href}%
\providecommand \@@href[1]{\endgroup#1\@@endlink}%
\providecommand \@sanitize@url [0]{\catcode `\\12\catcode `\$12\catcode
  `\&12\catcode `\#12\catcode `\^12\catcode `\_12\catcode `\%12\relax}%
\providecommand \@@startlink[1]{}%
\providecommand \@@endlink[0]{}%
\providecommand \url  [0]{\begingroup\@sanitize@url \@url }%
\providecommand \@url [1]{\endgroup\@href {#1}{\urlprefix }}%
\providecommand \urlprefix  [0]{URL }%
\providecommand \Eprint [0]{\href }%
\providecommand \doibase [0]{http://dx.doi.org/}%
\providecommand \selectlanguage [0]{\@gobble}%
\providecommand \bibinfo  [0]{\@secondoftwo}%
\providecommand \bibfield  [0]{\@secondoftwo}%
\providecommand \translation [1]{[#1]}%
\providecommand \BibitemOpen [0]{}%
\providecommand \bibitemStop [0]{}%
\providecommand \bibitemNoStop [0]{.\EOS\space}%
\providecommand \EOS [0]{\spacefactor3000\relax}%
\providecommand \BibitemShut  [1]{\csname bibitem#1\endcsname}%
\let\auto@bib@innerbib\@empty
\bibitem [{\citenamefont {Chakravarty}\ and\ \citenamefont
  {Hsu}(2015)}]{Chakravarty2015}%
  \BibitemOpen
  \bibfield  {author} {\bibinfo {author} {\bibfnamefont {S.}~\bibnamefont
  {Chakravarty}}\ and\ \bibinfo {author} {\bibfnamefont {C.-H.}\ \bibnamefont
  {Hsu}},\ }\href@noop {} {\bibfield  {journal} {\bibinfo  {journal} {Modern
  Physics Letters B}\ }\textbf {\bibinfo {volume} {29}},\ \bibinfo {pages}
  {1540053} (\bibinfo {year} {2015})}\BibitemShut {NoStop}%
\bibitem [{\citenamefont {Kallin}\ and\ \citenamefont
  {Berlinsky}(2016)}]{Kallin2016}%
  \BibitemOpen
  \bibfield  {author} {\bibinfo {author} {\bibfnamefont {C.}~\bibnamefont
  {Kallin}}\ and\ \bibinfo {author} {\bibfnamefont {J.}~\bibnamefont
  {Berlinsky}},\ }\href@noop {} {\bibfield  {journal} {\bibinfo  {journal}
  {Reports on Progress in Physics}\ }\textbf {\bibinfo {volume} {79}},\
  \bibinfo {pages} {054502} (\bibinfo {year} {2016})}\BibitemShut {NoStop}%
\bibitem [{\citenamefont {Sato}\ and\ \citenamefont
  {Fujimoto}(2016)}]{Sato2016}%
  \BibitemOpen
  \bibfield  {author} {\bibinfo {author} {\bibfnamefont {M.}~\bibnamefont
  {Sato}}\ and\ \bibinfo {author} {\bibfnamefont {S.}~\bibnamefont
  {Fujimoto}},\ }\href@noop {} {\bibfield  {journal} {\bibinfo  {journal} {J.
  Phys. Soc. Jpn.}\ } (\bibinfo {year} {2016})}\BibitemShut {NoStop}%
\bibitem [{\citenamefont {Kozii}\ \emph {et~al.}(2016)\citenamefont {Kozii},
  \citenamefont {Venderbos},\ and\ \citenamefont {Fu}}]{Kozii2016}%
  \BibitemOpen
  \bibfield  {author} {\bibinfo {author} {\bibfnamefont {V.}~\bibnamefont
  {Kozii}}, \bibinfo {author} {\bibfnamefont {J.~W.~F.}\ \bibnamefont
  {Venderbos}}, \ and\ \bibinfo {author} {\bibfnamefont {L.}~\bibnamefont
  {Fu}},\ }\href {http://arxiv.org/abs/1607.08243} {\  (\bibinfo {year}
  {2016})},\ \Eprint {http://arxiv.org/abs/1607.08243} {arXiv:1607.08243}
  \BibitemShut {NoStop}%
\bibitem [{\citenamefont {Bauer}\ \emph {et~al.}(2002)\citenamefont {Bauer},
  \citenamefont {Frederick}, \citenamefont {Ho}, \citenamefont {Zapf},\ and\
  \citenamefont {Maple}}]{Bauer2002}%
  \BibitemOpen
  \bibfield  {author} {\bibinfo {author} {\bibfnamefont {E.~D.}\ \bibnamefont
  {Bauer}}, \bibinfo {author} {\bibfnamefont {N.~A.}\ \bibnamefont
  {Frederick}}, \bibinfo {author} {\bibfnamefont {P.-C.}\ \bibnamefont {Ho}},
  \bibinfo {author} {\bibfnamefont {V.~S.}\ \bibnamefont {Zapf}}, \ and\
  \bibinfo {author} {\bibfnamefont {M.~B.}\ \bibnamefont {Maple}},\ }\href
  {\doibase 10.1103/PhysRevB.65.100506} {\bibfield  {journal} {\bibinfo
  {journal} {Phys. Rev. B}\ }\textbf {\bibinfo {volume} {65}},\ \bibinfo
  {pages} {100506} (\bibinfo {year} {2002})}\BibitemShut {NoStop}%
\bibitem [{\citenamefont {Maple}\ \emph {et~al.}(2006)\citenamefont {Maple},
  \citenamefont {Frederick}, \citenamefont {Ho}, \citenamefont {Yuhasz},\ and\
  \citenamefont {Yanagisawa}}]{Maple2006}%
  \BibitemOpen
  \bibfield  {author} {\bibinfo {author} {\bibfnamefont {M.~B.}\ \bibnamefont
  {Maple}}, \bibinfo {author} {\bibfnamefont {N.~A.}\ \bibnamefont
  {Frederick}}, \bibinfo {author} {\bibfnamefont {P.-C.}\ \bibnamefont {Ho}},
  \bibinfo {author} {\bibfnamefont {W.~M.}\ \bibnamefont {Yuhasz}}, \ and\
  \bibinfo {author} {\bibfnamefont {T.}~\bibnamefont {Yanagisawa}},\ }\href
  {\doibase 10.1007/s10948-006-0165-8} {\bibfield  {journal} {\bibinfo
  {journal} {Journal of Superconductivity and Novel Magnetism}\ }\textbf
  {\bibinfo {volume} {19}},\ \bibinfo {pages} {299} (\bibinfo {year}
  {2006})}\BibitemShut {NoStop}%
\bibitem [{\citenamefont {Aoki}\ \emph {et~al.}(2002)\citenamefont {Aoki},
  \citenamefont {Namiki}, \citenamefont {Ohsaki}, \citenamefont {Saha},
  \citenamefont {Sugawara},\ and\ \citenamefont {Sato}}]{Aoki2002}%
  \BibitemOpen
  \bibfield  {author} {\bibinfo {author} {\bibfnamefont {Y.}~\bibnamefont
  {Aoki}}, \bibinfo {author} {\bibfnamefont {T.}~\bibnamefont {Namiki}},
  \bibinfo {author} {\bibfnamefont {S.}~\bibnamefont {Ohsaki}}, \bibinfo
  {author} {\bibfnamefont {S.~R.}\ \bibnamefont {Saha}}, \bibinfo {author}
  {\bibfnamefont {H.}~\bibnamefont {Sugawara}}, \ and\ \bibinfo {author}
  {\bibfnamefont {H.}~\bibnamefont {Sato}},\ }\href {\doibase
  10.1143/JPSJ.71.2098} {\bibfield  {journal} {\bibinfo  {journal} {Journal of
  the Physical Society of Japan}\ }\textbf {\bibinfo {volume} {71}},\ \bibinfo
  {pages} {2098} (\bibinfo {year} {2002})}\BibitemShut {NoStop}%
\bibitem [{\citenamefont {Izawa}\ \emph {et~al.}(2003)\citenamefont {Izawa},
  \citenamefont {Nakajima}, \citenamefont {Goryo}, \citenamefont {Matsuda},
  \citenamefont {Osaki}, \citenamefont {Sugawara}, \citenamefont {Sato},
  \citenamefont {Thalmeier},\ and\ \citenamefont {Maki}}]{Izawa2003}%
  \BibitemOpen
  \bibfield  {author} {\bibinfo {author} {\bibfnamefont {K.}~\bibnamefont
  {Izawa}}, \bibinfo {author} {\bibfnamefont {Y.}~\bibnamefont {Nakajima}},
  \bibinfo {author} {\bibfnamefont {J.}~\bibnamefont {Goryo}}, \bibinfo
  {author} {\bibfnamefont {Y.}~\bibnamefont {Matsuda}}, \bibinfo {author}
  {\bibfnamefont {S.}~\bibnamefont {Osaki}}, \bibinfo {author} {\bibfnamefont
  {H.}~\bibnamefont {Sugawara}}, \bibinfo {author} {\bibfnamefont
  {H.}~\bibnamefont {Sato}}, \bibinfo {author} {\bibfnamefont {P.}~\bibnamefont
  {Thalmeier}}, \ and\ \bibinfo {author} {\bibfnamefont {K.}~\bibnamefont
  {Maki}},\ }\href {\doibase 10.1103/PhysRevLett.90.117001} {\bibfield
  {journal} {\bibinfo  {journal} {Phys. Rev. Lett.}\ }\textbf {\bibinfo
  {volume} {90}},\ \bibinfo {pages} {117001} (\bibinfo {year}
  {2003})}\BibitemShut {NoStop}%
\bibitem [{\citenamefont {Vollmer}\ \emph {et~al.}(2003)\citenamefont
  {Vollmer}, \citenamefont {Fai\ss{}t}, \citenamefont {Pfleiderer},
  \citenamefont {v.~L\"ohneysen}, \citenamefont {Bauer}, \citenamefont {Ho},
  \citenamefont {Zapf},\ and\ \citenamefont {Maple}}]{Vollmer2003}%
  \BibitemOpen
  \bibfield  {author} {\bibinfo {author} {\bibfnamefont {R.}~\bibnamefont
  {Vollmer}}, \bibinfo {author} {\bibfnamefont {A.}~\bibnamefont {Fai\ss{}t}},
  \bibinfo {author} {\bibfnamefont {C.}~\bibnamefont {Pfleiderer}}, \bibinfo
  {author} {\bibfnamefont {H.}~\bibnamefont {v.~L\"ohneysen}}, \bibinfo
  {author} {\bibfnamefont {E.~D.}\ \bibnamefont {Bauer}}, \bibinfo {author}
  {\bibfnamefont {P.-C.}\ \bibnamefont {Ho}}, \bibinfo {author} {\bibfnamefont
  {V.}~\bibnamefont {Zapf}}, \ and\ \bibinfo {author} {\bibfnamefont {M.~B.}\
  \bibnamefont {Maple}},\ }\href {\doibase 10.1103/PhysRevLett.90.057001}
  {\bibfield  {journal} {\bibinfo  {journal} {Phys. Rev. Lett.}\ }\textbf
  {\bibinfo {volume} {90}},\ \bibinfo {pages} {057001} (\bibinfo {year}
  {2003})}\BibitemShut {NoStop}%
\bibitem [{\citenamefont {Aoki}\ \emph {et~al.}(2007)\citenamefont {Aoki},
  \citenamefont {Tayama}, \citenamefont {Sakakibara}, \citenamefont {Kuwahara},
  \citenamefont {Iwasa}, \citenamefont {Kohgi}, \citenamefont {Higemoto},
  \citenamefont {MacLaughlin}, \citenamefont {Sugawara},\ and\ \citenamefont
  {Sato}}]{Aoki2007}%
  \BibitemOpen
  \bibfield  {author} {\bibinfo {author} {\bibfnamefont {Y.}~\bibnamefont
  {Aoki}}, \bibinfo {author} {\bibfnamefont {T.}~\bibnamefont {Tayama}},
  \bibinfo {author} {\bibfnamefont {T.}~\bibnamefont {Sakakibara}}, \bibinfo
  {author} {\bibfnamefont {K.}~\bibnamefont {Kuwahara}}, \bibinfo {author}
  {\bibfnamefont {K.}~\bibnamefont {Iwasa}}, \bibinfo {author} {\bibfnamefont
  {M.}~\bibnamefont {Kohgi}}, \bibinfo {author} {\bibfnamefont
  {W.}~\bibnamefont {Higemoto}}, \bibinfo {author} {\bibfnamefont {D.~E.}\
  \bibnamefont {MacLaughlin}}, \bibinfo {author} {\bibfnamefont
  {H.}~\bibnamefont {Sugawara}}, \ and\ \bibinfo {author} {\bibfnamefont
  {H.}~\bibnamefont {Sato}},\ }\href {\doibase 10.1143/JPSJ.76.051006}
  {\bibfield  {journal} {\bibinfo  {journal} {Journal of the Physical Society
  of Japan}\ }\textbf {\bibinfo {volume} {76}},\ \bibinfo {pages} {051006}
  (\bibinfo {year} {2007})}\BibitemShut {NoStop}%
\bibitem [{\citenamefont {Miyake}\ \emph {et~al.}(2003)\citenamefont {Miyake},
  \citenamefont {Kohno},\ and\ \citenamefont {Harima}}]{Miyake2003}%
  \BibitemOpen
  \bibfield  {author} {\bibinfo {author} {\bibfnamefont {K.}~\bibnamefont
  {Miyake}}, \bibinfo {author} {\bibfnamefont {H.}~\bibnamefont {Kohno}}, \
  and\ \bibinfo {author} {\bibfnamefont {H.}~\bibnamefont {Harima}},\ }\href
  {http://stacks.iop.org/0953-8984/15/i=19/a=102} {\bibfield  {journal}
  {\bibinfo  {journal} {Journal of Physics: Condensed Matter}\ }\textbf
  {\bibinfo {volume} {15}},\ \bibinfo {pages} {L275} (\bibinfo {year}
  {2003})}\BibitemShut {NoStop}%
\bibitem [{\citenamefont {Aoki}\ \emph {et~al.}(2003)\citenamefont {Aoki},
  \citenamefont {Tsuchiya}, \citenamefont {Kanayama}, \citenamefont {Saha},
  \citenamefont {Sugawara}, \citenamefont {Sato}, \citenamefont {Higemoto},
  \citenamefont {Koda}, \citenamefont {Ohishi}, \citenamefont {Nishiyama},\
  and\ \citenamefont {Kadono}}]{Aoki2003}%
  \BibitemOpen
  \bibfield  {author} {\bibinfo {author} {\bibfnamefont {Y.}~\bibnamefont
  {Aoki}}, \bibinfo {author} {\bibfnamefont {A.}~\bibnamefont {Tsuchiya}},
  \bibinfo {author} {\bibfnamefont {T.}~\bibnamefont {Kanayama}}, \bibinfo
  {author} {\bibfnamefont {S.~R.}\ \bibnamefont {Saha}}, \bibinfo {author}
  {\bibfnamefont {H.}~\bibnamefont {Sugawara}}, \bibinfo {author}
  {\bibfnamefont {H.}~\bibnamefont {Sato}}, \bibinfo {author} {\bibfnamefont
  {W.}~\bibnamefont {Higemoto}}, \bibinfo {author} {\bibfnamefont
  {A.}~\bibnamefont {Koda}}, \bibinfo {author} {\bibfnamefont {K.}~\bibnamefont
  {Ohishi}}, \bibinfo {author} {\bibfnamefont {K.}~\bibnamefont {Nishiyama}}, \
  and\ \bibinfo {author} {\bibfnamefont {R.}~\bibnamefont {Kadono}},\ }\href
  {\doibase 10.1103/PhysRevLett.91.067003} {\bibfield  {journal} {\bibinfo
  {journal} {Phys. Rev. Lett.}\ }\textbf {\bibinfo {volume} {91}},\ \bibinfo
  {pages} {067003} (\bibinfo {year} {2003})}\BibitemShut {NoStop}%
\bibitem [{\citenamefont {M{\'e}asson}\ \emph {et~al.}(2006)\citenamefont
  {M{\'e}asson}, \citenamefont {Braithwaite}, \citenamefont {Salce},
  \citenamefont {Flouquet}, \citenamefont {Lapertot}, \citenamefont
  {P{\'e}caut}, \citenamefont {Seyfarth}, \citenamefont {Brison}, \citenamefont
  {Sugawara},\ and\ \citenamefont {Sato}}]{Méasson2006}%
  \BibitemOpen
  \bibfield  {author} {\bibinfo {author} {\bibfnamefont {M.-A.}\ \bibnamefont
  {M{\'e}asson}}, \bibinfo {author} {\bibfnamefont {D.}~\bibnamefont
  {Braithwaite}}, \bibinfo {author} {\bibfnamefont {B.}~\bibnamefont {Salce}},
  \bibinfo {author} {\bibfnamefont {J.}~\bibnamefont {Flouquet}}, \bibinfo
  {author} {\bibfnamefont {G.}~\bibnamefont {Lapertot}}, \bibinfo {author}
  {\bibfnamefont {J.}~\bibnamefont {P{\'e}caut}}, \bibinfo {author}
  {\bibfnamefont {G.}~\bibnamefont {Seyfarth}}, \bibinfo {author}
  {\bibfnamefont {J.-P.}\ \bibnamefont {Brison}}, \bibinfo {author}
  {\bibfnamefont {H.}~\bibnamefont {Sugawara}}, \ and\ \bibinfo {author}
  {\bibfnamefont {H.}~\bibnamefont {Sato}},\ }\href {\doibase
  http://dx.doi.org/10.1016/j.physb.2006.01.169} {\bibfield  {journal}
  {\bibinfo  {journal} {Physica B: Condensed Matter}\ }\textbf {\bibinfo
  {volume} {378–380}},\ \bibinfo {pages} {56 } (\bibinfo {year} {2006})},\
  \bibinfo {note} {proceedings of the International Conference on Strongly
  Correlated Electron Systems}\BibitemShut {NoStop}%
\bibitem [{\citenamefont {Seyfarth}\ \emph {et~al.}(2006)\citenamefont
  {Seyfarth}, \citenamefont {Brison}, \citenamefont {M\'easson}, \citenamefont
  {Braithwaite}, \citenamefont {Lapertot},\ and\ \citenamefont
  {Flouquet}}]{Seyfarth2006}%
  \BibitemOpen
  \bibfield  {author} {\bibinfo {author} {\bibfnamefont {G.}~\bibnamefont
  {Seyfarth}}, \bibinfo {author} {\bibfnamefont {J.~P.}\ \bibnamefont
  {Brison}}, \bibinfo {author} {\bibfnamefont {M.-A.}\ \bibnamefont
  {M\'easson}}, \bibinfo {author} {\bibfnamefont {D.}~\bibnamefont
  {Braithwaite}}, \bibinfo {author} {\bibfnamefont {G.}~\bibnamefont
  {Lapertot}}, \ and\ \bibinfo {author} {\bibfnamefont {J.}~\bibnamefont
  {Flouquet}},\ }\href {\doibase 10.1103/PhysRevLett.97.236403} {\bibfield
  {journal} {\bibinfo  {journal} {Phys. Rev. Lett.}\ }\textbf {\bibinfo
  {volume} {97}},\ \bibinfo {pages} {236403} (\bibinfo {year}
  {2006})}\BibitemShut {NoStop}%
\bibitem [{\citenamefont {M\'easson}\ \emph {et~al.}(2008)\citenamefont
  {M\'easson}, \citenamefont {Braithwaite}, \citenamefont {Lapertot},
  \citenamefont {Brison}, \citenamefont {Flouquet}, \citenamefont {Bordet},
  \citenamefont {Sugawara},\ and\ \citenamefont {Canfield}}]{Measson2008}%
  \BibitemOpen
  \bibfield  {author} {\bibinfo {author} {\bibfnamefont {M.-A.}\ \bibnamefont
  {M\'easson}}, \bibinfo {author} {\bibfnamefont {D.}~\bibnamefont
  {Braithwaite}}, \bibinfo {author} {\bibfnamefont {G.}~\bibnamefont
  {Lapertot}}, \bibinfo {author} {\bibfnamefont {J.-P.}\ \bibnamefont
  {Brison}}, \bibinfo {author} {\bibfnamefont {J.}~\bibnamefont {Flouquet}},
  \bibinfo {author} {\bibfnamefont {P.}~\bibnamefont {Bordet}}, \bibinfo
  {author} {\bibfnamefont {H.}~\bibnamefont {Sugawara}}, \ and\ \bibinfo
  {author} {\bibfnamefont {P.~C.}\ \bibnamefont {Canfield}},\ }\href {\doibase
  10.1103/PhysRevB.77.134517} {\bibfield  {journal} {\bibinfo  {journal} {Phys.
  Rev. B}\ }\textbf {\bibinfo {volume} {77}},\ \bibinfo {pages} {134517}
  (\bibinfo {year} {2008})}\BibitemShut {NoStop}%
\bibitem [{\citenamefont {Andraka}\ and\ \citenamefont
  {Pocsy}(2012)}]{Andraka2012}%
  \BibitemOpen
  \bibfield  {author} {\bibinfo {author} {\bibfnamefont {B.}~\bibnamefont
  {Andraka}}\ and\ \bibinfo {author} {\bibfnamefont {K.}~\bibnamefont
  {Pocsy}},\ }\href {\doibase 10.1063/1.3672061} {\bibfield  {journal}
  {\bibinfo  {journal} {Journal of Applied Physics}\ }\textbf {\bibinfo
  {volume} {111}},\ \bibinfo {pages} {07E115} (\bibinfo {year}
  {2012})}\BibitemShut {NoStop}%
\bibitem [{\citenamefont {McBriarty}\ \emph {et~al.}(2009)\citenamefont
  {McBriarty}, \citenamefont {Kumar}, \citenamefont {Stewart},\ and\
  \citenamefont {Andraka}}]{McBriarty2009}%
  \BibitemOpen
  \bibfield  {author} {\bibinfo {author} {\bibfnamefont {M.~E.}\ \bibnamefont
  {McBriarty}}, \bibinfo {author} {\bibfnamefont {P.}~\bibnamefont {Kumar}},
  \bibinfo {author} {\bibfnamefont {G.~R.}\ \bibnamefont {Stewart}}, \ and\
  \bibinfo {author} {\bibfnamefont {B.}~\bibnamefont {Andraka}},\ }\href
  {\doibase 10.1088/0953-8984/21/38/385701} {\bibfield  {journal} {\bibinfo
  {journal} {Journal of Physics: Condensed Matter}\ }\textbf {\bibinfo {volume}
  {21}},\ \bibinfo {pages} {385701} (\bibinfo {year} {2009})}\BibitemShut
  {NoStop}%
\bibitem [{\citenamefont {Sauls}(1994)}]{Sauls1994}%
  \BibitemOpen
  \bibfield  {author} {\bibinfo {author} {\bibfnamefont {J.}~\bibnamefont
  {Sauls}},\ }\href@noop {} {\bibfield  {journal} {\bibinfo  {journal}
  {Advances in Physics}\ }\textbf {\bibinfo {volume} {43}},\ \bibinfo {pages}
  {113} (\bibinfo {year} {1994})}\BibitemShut {NoStop}%
\bibitem [{Ord()}]{Orderofmagnitude}%
  \BibitemOpen
  \href@noop {} {}\bibinfo {note} {Currently the two known scenarios that
  result in a finite Kerr effect in a TRSB superconductor require either
  Magnetic scattering or interband coupling. See discussion in
  \cite{Kapitulnik2015} and references therein.}\BibitemShut {Stop}%
\bibitem [{\citenamefont {Xia}\ \emph {et~al.}(2006{\natexlab{a}})\citenamefont
  {Xia}, \citenamefont {Maeno}, \citenamefont {Beyersdorf}, \citenamefont
  {Fejer},\ and\ \citenamefont {Kapitulnik}}]{Xia2006}%
  \BibitemOpen
  \bibfield  {author} {\bibinfo {author} {\bibfnamefont {J.}~\bibnamefont
  {Xia}}, \bibinfo {author} {\bibfnamefont {Y.}~\bibnamefont {Maeno}}, \bibinfo
  {author} {\bibfnamefont {P.~T.}\ \bibnamefont {Beyersdorf}}, \bibinfo
  {author} {\bibfnamefont {M.~M.}\ \bibnamefont {Fejer}}, \ and\ \bibinfo
  {author} {\bibfnamefont {A.}~\bibnamefont {Kapitulnik}},\ }\href {\doibase
  10.1103/PhysRevLett.97.167002} {\bibfield  {journal} {\bibinfo  {journal}
  {Phys. Rev. Lett.}\ }\textbf {\bibinfo {volume} {97}},\ \bibinfo {pages}
  {167002} (\bibinfo {year} {2006}{\natexlab{a}})}\BibitemShut {NoStop}%
\bibitem [{\citenamefont {Kapitulnik}\ \emph
  {et~al.}(2009{\natexlab{a}})\citenamefont {Kapitulnik}, \citenamefont {Xia},\
  and\ \citenamefont {Schemm}}]{Kapitulnik2009}%
  \BibitemOpen
  \bibfield  {author} {\bibinfo {author} {\bibfnamefont {A.}~\bibnamefont
  {Kapitulnik}}, \bibinfo {author} {\bibfnamefont {J.}~\bibnamefont {Xia}}, \
  and\ \bibinfo {author} {\bibfnamefont {E.}~\bibnamefont {Schemm}},\
  }\href@noop {} {\bibfield  {journal} {\bibinfo  {journal} {Physica B:
  Condensed Matter}\ }\textbf {\bibinfo {volume} {404}},\ \bibinfo {pages}
  {507} (\bibinfo {year} {2009}{\natexlab{a}})}\BibitemShut {NoStop}%
\bibitem [{\citenamefont {Kapitulnik}(2015)}]{Kapitulnik2015}%
  \BibitemOpen
  \bibfield  {author} {\bibinfo {author} {\bibfnamefont {A.}~\bibnamefont
  {Kapitulnik}},\ }\href {\doibase 10.1016/j.physb.2014.11.059} {\bibfield
  {journal} {\bibinfo  {journal} {Physica B-Condensed Matter}\ }\textbf
  {\bibinfo {volume} {460}},\ \bibinfo {pages} {151} (\bibinfo {year}
  {2015})}\BibitemShut {NoStop}%
\bibitem [{\citenamefont {Schemm}\ \emph {et~al.}(2015)\citenamefont {Schemm},
  \citenamefont {Baumbach}, \citenamefont {Tobash}, \citenamefont {Ronning},
  \citenamefont {Bauer},\ and\ \citenamefont {Kapitulnik}}]{Schemm2015}%
  \BibitemOpen
  \bibfield  {author} {\bibinfo {author} {\bibfnamefont {E.~R.}\ \bibnamefont
  {Schemm}}, \bibinfo {author} {\bibfnamefont {R.~E.}\ \bibnamefont
  {Baumbach}}, \bibinfo {author} {\bibfnamefont {P.~H.}\ \bibnamefont
  {Tobash}}, \bibinfo {author} {\bibfnamefont {F.}~\bibnamefont {Ronning}},
  \bibinfo {author} {\bibfnamefont {E.~D.}\ \bibnamefont {Bauer}}, \ and\
  \bibinfo {author} {\bibfnamefont {A.}~\bibnamefont {Kapitulnik}},\
  }\href@noop {} {\bibfield  {journal} {\bibinfo  {journal} {Phys. Rev. B}\
  }\textbf {\bibinfo {volume} {91}},\ \bibinfo {pages} {140506} (\bibinfo
  {year} {2015})}\BibitemShut {NoStop}%
\bibitem [{\citenamefont {Schemm}\ \emph {et~al.}(2014)\citenamefont {Schemm},
  \citenamefont {Gannon}, \citenamefont {Wishne}, \citenamefont {Halperin},\
  and\ \citenamefont {Kapitulnik}}]{Schemm2014}%
  \BibitemOpen
  \bibfield  {author} {\bibinfo {author} {\bibfnamefont {E.~R.}\ \bibnamefont
  {Schemm}}, \bibinfo {author} {\bibfnamefont {W.~J.}\ \bibnamefont {Gannon}},
  \bibinfo {author} {\bibfnamefont {C.~M.}\ \bibnamefont {Wishne}}, \bibinfo
  {author} {\bibfnamefont {W.~P.}\ \bibnamefont {Halperin}}, \ and\ \bibinfo
  {author} {\bibfnamefont {A.}~\bibnamefont {Kapitulnik}},\ }\href@noop {}
  {\bibfield  {journal} {\bibinfo  {journal} {Science}\ }\textbf {\bibinfo
  {volume} {345}},\ \bibinfo {pages} {190} (\bibinfo {year}
  {2014})}\BibitemShut {NoStop}%
\bibitem [{\citenamefont {Schemm}\ \emph {et~al.}(2017)\citenamefont {Schemm},
  \citenamefont {Levenson-Falk},\ and\ \citenamefont
  {Kapitulnik}}]{Schemm2017}%
  \BibitemOpen
  \bibfield  {author} {\bibinfo {author} {\bibfnamefont {E.}~\bibnamefont
  {Schemm}}, \bibinfo {author} {\bibfnamefont {E.}~\bibnamefont
  {Levenson-Falk}}, \ and\ \bibinfo {author} {\bibfnamefont {A.}~\bibnamefont
  {Kapitulnik}},\ }\href {\doibase https://doi.org/10.1016/j.physc.2016.11.012}
  {\bibfield  {journal} {\bibinfo  {journal} {Physica C: Superconductivity and
  its Applications}\ }\textbf {\bibinfo {volume} {535}},\ \bibinfo {pages} {13
  } (\bibinfo {year} {2017})}\BibitemShut {NoStop}%
\bibitem [{\citenamefont {Xia}\ \emph {et~al.}(2006{\natexlab{b}})\citenamefont
  {Xia}, \citenamefont {Beyersdorf}, \citenamefont {Fejer},\ and\ \citenamefont
  {Kapitulnik}}]{Xia2006a}%
  \BibitemOpen
  \bibfield  {author} {\bibinfo {author} {\bibfnamefont {J.}~\bibnamefont
  {Xia}}, \bibinfo {author} {\bibfnamefont {P.~T.}\ \bibnamefont {Beyersdorf}},
  \bibinfo {author} {\bibfnamefont {M.~M.}\ \bibnamefont {Fejer}}, \ and\
  \bibinfo {author} {\bibfnamefont {A.}~\bibnamefont {Kapitulnik}},\
  }\href@noop {} {\bibfield  {journal} {\bibinfo  {journal} {Appl. Phys.
  Lett.}\ }\textbf {\bibinfo {volume} {89}} (\bibinfo {year}
  {2006}{\natexlab{b}})}\BibitemShut {NoStop}%
\bibitem [{\citenamefont {Kapitulnik}\ \emph
  {et~al.}(2009{\natexlab{b}})\citenamefont {Kapitulnik}, \citenamefont {Xia},
  \citenamefont {Schemm},\ and\ \citenamefont {Palevski}}]{Kapitulnik2009a}%
  \BibitemOpen
  \bibfield  {author} {\bibinfo {author} {\bibfnamefont {A.}~\bibnamefont
  {Kapitulnik}}, \bibinfo {author} {\bibfnamefont {J.}~\bibnamefont {Xia}},
  \bibinfo {author} {\bibfnamefont {E.}~\bibnamefont {Schemm}}, \ and\ \bibinfo
  {author} {\bibfnamefont {A.}~\bibnamefont {Palevski}},\ }\href@noop {}
  {\bibfield  {journal} {\bibinfo  {journal} {New Journal of Physics}\ }\textbf
  {\bibinfo {volume} {11}} (\bibinfo {year} {2009}{\natexlab{b}})}\BibitemShut
  {NoStop}%
\bibitem [{\citenamefont {Bauer}\ \emph {et~al.}(2001)\citenamefont {Bauer},
  \citenamefont {Slebarski}, \citenamefont {Freeman}, \citenamefont {Sirvent},\
  and\ \citenamefont {Maple}}]{Bauer2001}%
  \BibitemOpen
  \bibfield  {author} {\bibinfo {author} {\bibfnamefont {E.~D.}\ \bibnamefont
  {Bauer}}, \bibinfo {author} {\bibfnamefont {A.}~\bibnamefont {Slebarski}},
  \bibinfo {author} {\bibfnamefont {E.~J.}\ \bibnamefont {Freeman}}, \bibinfo
  {author} {\bibfnamefont {C.}~\bibnamefont {Sirvent}}, \ and\ \bibinfo
  {author} {\bibfnamefont {M.~B.}\ \bibnamefont {Maple}},\ }\href
  {http://stacks.iop.org/0953-8984/13/i=20/a=310} {\bibfield  {journal}
  {\bibinfo  {journal} {Journal of Physics: Condensed Matter}\ }\textbf
  {\bibinfo {volume} {13}},\ \bibinfo {pages} {4495} (\bibinfo {year}
  {2001})}\BibitemShut {NoStop}%
\bibitem [{\citenamefont {Braun}\ and\ \citenamefont
  {Jeitschko}(1980)}]{Braun1980}%
  \BibitemOpen
  \bibfield  {author} {\bibinfo {author} {\bibfnamefont {D.}~\bibnamefont
  {Braun}}\ and\ \bibinfo {author} {\bibfnamefont {W.}~\bibnamefont
  {Jeitschko}},\ }\href@noop {} {\bibfield  {journal} {\bibinfo  {journal}
  {Journal of the Less Common Metals}\ }\textbf {\bibinfo {volume} {72}},\
  \bibinfo {pages} {147} (\bibinfo {year} {1980})}\BibitemShut {NoStop}%
\bibitem [{\citenamefont {Frederick}\ and\ \citenamefont
  {Maple}(2003)}]{Frederick2003}%
  \BibitemOpen
  \bibfield  {author} {\bibinfo {author} {\bibfnamefont {N.~A.}\ \bibnamefont
  {Frederick}}\ and\ \bibinfo {author} {\bibfnamefont {M.~B.}\ \bibnamefont
  {Maple}},\ }\href {http://stacks.iop.org/0953-8984/15/i=27/a=310} {\bibfield
  {journal} {\bibinfo  {journal} {Journal of Physics: Condensed Matter}\
  }\textbf {\bibinfo {volume} {15}},\ \bibinfo {pages} {4789} (\bibinfo {year}
  {2003})}\BibitemShut {NoStop}%
\bibitem [{\citenamefont {Brandt}(1998)}]{Brandt1998}%
  \BibitemOpen
  \bibfield  {author} {\bibinfo {author} {\bibfnamefont {E.~H.}\ \bibnamefont
  {Brandt}},\ }\href {\doibase 10.1103/PhysRevB.58.6523} {\bibfield  {journal}
  {\bibinfo  {journal} {Phys. Rev. B}\ }\textbf {\bibinfo {volume} {58}},\
  \bibinfo {pages} {6523} (\bibinfo {year} {1998})}\BibitemShut {NoStop}%
\bibitem [{\citenamefont {Khasanov}\ \emph {et~al.}(2007)\citenamefont
  {Khasanov}, \citenamefont {Str\"assle}, \citenamefont {Di~Castro},
  \citenamefont {Masui}, \citenamefont {Miyasaka}, \citenamefont {Tajima},
  \citenamefont {Bussmann-Holder},\ and\ \citenamefont
  {Keller}}]{Khasanov2007}%
  \BibitemOpen
  \bibfield  {author} {\bibinfo {author} {\bibfnamefont {R.}~\bibnamefont
  {Khasanov}}, \bibinfo {author} {\bibfnamefont {S.}~\bibnamefont
  {Str\"assle}}, \bibinfo {author} {\bibfnamefont {D.}~\bibnamefont
  {Di~Castro}}, \bibinfo {author} {\bibfnamefont {T.}~\bibnamefont {Masui}},
  \bibinfo {author} {\bibfnamefont {S.}~\bibnamefont {Miyasaka}}, \bibinfo
  {author} {\bibfnamefont {S.}~\bibnamefont {Tajima}}, \bibinfo {author}
  {\bibfnamefont {A.}~\bibnamefont {Bussmann-Holder}}, \ and\ \bibinfo {author}
  {\bibfnamefont {H.}~\bibnamefont {Keller}},\ }\href {\doibase
  10.1103/PhysRevLett.99.237601} {\bibfield  {journal} {\bibinfo  {journal}
  {Phys. Rev. Lett.}\ }\textbf {\bibinfo {volume} {99}},\ \bibinfo {pages}
  {237601} (\bibinfo {year} {2007})}\BibitemShut {NoStop}%
\bibitem [{\citenamefont {Cichorek}\ \emph {et~al.}(2005)\citenamefont
  {Cichorek}, \citenamefont {Mota}, \citenamefont {Steglich}, \citenamefont
  {Frederick}, \citenamefont {Yuhasz},\ and\ \citenamefont
  {Maple}}]{Cichorek2005}%
  \BibitemOpen
  \bibfield  {author} {\bibinfo {author} {\bibfnamefont {T.}~\bibnamefont
  {Cichorek}}, \bibinfo {author} {\bibfnamefont {A.~C.}\ \bibnamefont {Mota}},
  \bibinfo {author} {\bibfnamefont {F.}~\bibnamefont {Steglich}}, \bibinfo
  {author} {\bibfnamefont {N.~A.}\ \bibnamefont {Frederick}}, \bibinfo {author}
  {\bibfnamefont {W.~M.}\ \bibnamefont {Yuhasz}}, \ and\ \bibinfo {author}
  {\bibfnamefont {M.~B.}\ \bibnamefont {Maple}},\ }\href {\doibase
  10.1103/PhysRevLett.94.107002} {\bibfield  {journal} {\bibinfo  {journal}
  {Phys. Rev. Lett.}\ }\textbf {\bibinfo {volume} {94}},\ \bibinfo {pages}
  {107002} (\bibinfo {year} {2005})}\BibitemShut {NoStop}%
\bibitem [{sup()}]{supplement}%
  \BibitemOpen
  \href@noop {} {}\bibinfo {note} {See supplementary material, which includes
  Refs.~\cite{multifit,York2004}}\BibitemShut {NoStop}%
\bibitem [{sam()}]{sampleBD}%
  \BibitemOpen
  \href@noop {} {}\bibinfo {note} {Samples B and D were measured only at high
  power (38 $\mu$W). Both showed TRSB, but cannot be included in this analysis
  due to the absence of 20 $\mu$W data. Data on these samples is included in
  the Supplementary Material.}\BibitemShut {Stop}%
\bibitem [{\citenamefont {de~Lozanne}(1999)}]{deLozanneSPM}%
  \BibitemOpen
  \bibfield  {author} {\bibinfo {author} {\bibfnamefont {A.}~\bibnamefont
  {de~Lozanne}},\ }\href {http://stacks.iop.org/0953-2048/12/i=4/a=001}
  {\bibfield  {journal} {\bibinfo  {journal} {Superconductor Science and
  Technology}\ }\textbf {\bibinfo {volume} {12}},\ \bibinfo {pages} {R43}
  (\bibinfo {year} {1999})}\BibitemShut {NoStop}%
\bibitem [{\citenamefont {Gardner}\ \emph {et~al.}(2001)\citenamefont
  {Gardner}, \citenamefont {Wynn}, \citenamefont {Björnsson}, \citenamefont
  {Straver}, \citenamefont {Moler}, \citenamefont {Kirtley},\ and\
  \citenamefont {Ketchen}}]{MolerSQUID}%
  \BibitemOpen
  \bibfield  {author} {\bibinfo {author} {\bibfnamefont {B.~W.}\ \bibnamefont
  {Gardner}}, \bibinfo {author} {\bibfnamefont {J.~C.}\ \bibnamefont {Wynn}},
  \bibinfo {author} {\bibfnamefont {P.~G.}\ \bibnamefont {Björnsson}},
  \bibinfo {author} {\bibfnamefont {E.~W.~J.}\ \bibnamefont {Straver}},
  \bibinfo {author} {\bibfnamefont {K.~A.}\ \bibnamefont {Moler}}, \bibinfo
  {author} {\bibfnamefont {J.~R.}\ \bibnamefont {Kirtley}}, \ and\ \bibinfo
  {author} {\bibfnamefont {M.~B.}\ \bibnamefont {Ketchen}},\ }\href {\doibase
  10.1063/1.1364668} {\bibfield  {journal} {\bibinfo  {journal} {Review of
  Scientific Instruments}\ }\textbf {\bibinfo {volume} {72}},\ \bibinfo {pages}
  {2361} (\bibinfo {year} {2001})},\ \Eprint
  {http://arxiv.org/abs/http://dx.doi.org/10.1063/1.1364668}
  {http://dx.doi.org/10.1063/1.1364668} \BibitemShut {NoStop}%
\bibitem [{mul()}]{multifit}%
  \BibitemOpen
  \href@noop {} {}\bibinfo {note} {See
  https://www.mathworks.com/matlabcentral/fileexchange/40613-multiple-curve-fitting-with-common-parameters-using-nlinfit}\BibitemShut
  {NoStop}%
\bibitem [{\citenamefont {York}\ \emph {et~al.}(2004)\citenamefont {York},
  \citenamefont {Evensen}, \citenamefont {Martı́nez},\ and\ \citenamefont
  {Delgado}}]{York2004}%
  \BibitemOpen
  \bibfield  {author} {\bibinfo {author} {\bibfnamefont {D.}~\bibnamefont
  {York}}, \bibinfo {author} {\bibfnamefont {N.~M.}\ \bibnamefont {Evensen}},
  \bibinfo {author} {\bibfnamefont {M.~L.}\ \bibnamefont {Martı́nez}}, \ and\
  \bibinfo {author} {\bibfnamefont {J.~D.~B.}\ \bibnamefont {Delgado}},\ }\href
  {\doibase 10.1119/1.1632486} {\bibfield  {journal} {\bibinfo  {journal}
  {American Journal of Physics}\ }\textbf {\bibinfo {volume} {72}},\ \bibinfo
  {pages} {367} (\bibinfo {year} {2004})},\ \Eprint
  {http://arxiv.org/abs/https://doi.org/10.1119/1.1632486}
  {https://doi.org/10.1119/1.1632486} \BibitemShut {NoStop}%
\end{thebibliography}%


\begin{thebibliography}{2}%
\makeatletter
\providecommand \@ifxundefined [1]{%
 \@ifx{#1\undefined}
}%
\providecommand \@ifnum [1]{%
 \ifnum #1\expandafter \@firstoftwo
 \else \expandafter \@secondoftwo
 \fi
}%
\providecommand \@ifx [1]{%
 \ifx #1\expandafter \@firstoftwo
 \else \expandafter \@secondoftwo
 \fi
}%
\providecommand \natexlab [1]{#1}%
\providecommand \enquote  [1]{``#1''}%
\providecommand \bibnamefont  [1]{#1}%
\providecommand \bibfnamefont [1]{#1}%
\providecommand \citenamefont [1]{#1}%
\providecommand \href@noop [0]{\@secondoftwo}%
\providecommand \href [0]{\begingroup \@sanitize@url \@href}%
\providecommand \@href[1]{\@@startlink{#1}\@@href}%
\providecommand \@@href[1]{\endgroup#1\@@endlink}%
\providecommand \@sanitize@url [0]{\catcode `\\12\catcode `\$12\catcode
  `\&12\catcode `\#12\catcode `\^12\catcode `\_12\catcode `\%12\relax}%
\providecommand \@@startlink[1]{}%
\providecommand \@@endlink[0]{}%
\providecommand \url  [0]{\begingroup\@sanitize@url \@url }%
\providecommand \@url [1]{\endgroup\@href {#1}{\urlprefix }}%
\providecommand \urlprefix  [0]{URL }%
\providecommand \Eprint [0]{\href }%
\providecommand \doibase [0]{http://dx.doi.org/}%
\providecommand \selectlanguage [0]{\@gobble}%
\providecommand \bibinfo  [0]{\@secondoftwo}%
\providecommand \bibfield  [0]{\@secondoftwo}%
\providecommand \translation [1]{[#1]}%
\providecommand \BibitemOpen [0]{}%
\providecommand \bibitemStop [0]{}%
\providecommand \bibitemNoStop [0]{.\EOS\space}%
\providecommand \EOS [0]{\spacefactor3000\relax}%
\providecommand \BibitemShut  [1]{\csname bibitem#1\endcsname}%
\let\auto@bib@innerbib\@empty
\bibitem [{mul()}]{multifit}%
  \BibitemOpen
  \href@noop {} {}\bibinfo {note} {See
  https://www.mathworks.com/matlabcentral/fileexchange/40613-multiple-curve-fitting-with-common-parameters-using-nlinfit}\BibitemShut
  {NoStop}%
\bibitem [{\citenamefont {York}\ \emph {et~al.}(2004)\citenamefont {York},
  \citenamefont {Evensen}, \citenamefont {Martı́nez},\ and\ \citenamefont
  {Delgado}}]{York2004}%
  \BibitemOpen
  \bibfield  {author} {\bibinfo {author} {\bibfnamefont {D.}~\bibnamefont
  {York}}, \bibinfo {author} {\bibfnamefont {N.~M.}\ \bibnamefont {Evensen}},
  \bibinfo {author} {\bibfnamefont {M.~L.}\ \bibnamefont {Martı́nez}}, \ and\
  \bibinfo {author} {\bibfnamefont {J.~D.~B.}\ \bibnamefont {Delgado}},\ }\href
  {\doibase 10.1119/1.1632486} {\bibfield  {journal} {\bibinfo  {journal}
  {American Journal of Physics}\ }\textbf {\bibinfo {volume} {72}},\ \bibinfo
  {pages} {367} (\bibinfo {year} {2004})},\ \Eprint
  {http://arxiv.org/abs/https://doi.org/10.1119/1.1632486}
  {https://doi.org/10.1119/1.1632486} \BibitemShut {NoStop}%
\end{thebibliography}%
	
\end{document}


\title{Supplement to ``Polar Kerr Effect from Time Reversal Symmetry Breaking in the Heavy Fermion Superconductor PrOs$_4$Sb$_{12}$''}
	
	\author{E. M. Levenson-Falk}
	\email[Corresponding author: ]{elevenso@usc.edu}
	\affiliation{Department of Applied Physics, Stanford University, Stanford, CA 94305}
	\affiliation{Geballe Laboratory for Advanced Materials, Stanford University, Stanford, CA 94305}
	\affiliation{Department of Physics and Astronomy, University of Southern California, Los Angeles, CA 90089}
	\author{E. R. Schemm}
	\affiliation{Geballe Laboratory for Advanced Materials, Stanford University, Stanford, CA 94305}
	\author{Y. Aoki}
	\affiliation{Department of Physics, Tokyo Metropolitan University, Tokyo 192-0397}
	\author{M. B. Maple}
	\affiliation{Department of Physics and Center for Advanced Nanoscience, University of California San Diego, La Jolla, CA 92093}
	\author{A. Kapitulnik}
	\affiliation{Department of Applied Physics, Stanford University, Stanford, CA 94305}
	\affiliation{Department of Physics, Stanford University, Stanford, CA 94305}
	\affiliation{Stanford Institute for Materials and Energy Sciences, SLAC National Accelerator Laboratory, 2575 Sand Hill Road, Menlo Park, California 94025, USA}
	
	\maketitle

\section{Data Analysis and Fitting}
Kerr angle data was acquired at a rate of 1 sample per second. Due to the slow time constant of the RF lockin amplifier used to measure the signal, each data point is correlated with its $\sim8$ closest neighbors. In order to accurately compute error bars, we eliminated these correlations by dividing the data into 50-point ``chunks" and averaging the points in each chunk, leaving us with a set of averages $\{x_i\}$. These averages are then almost completely uncorrelated and thus represent independent samples with 50 s of averaging time. Before plotting the data vs temperature, the chunk averages are then binned into temperature bins, with all the $x_i$ in each bin averaged together. The points plotted in all $\theta_K(T)$ graphs are these bin averages. The error bars given are the 1-$\sigma$ standard error of the mean, computed as the standard deviation of the $x_i$ in each bin divided by the square root of the number of chunks.

Fits were performed using Matlab's built-in fitting procedures and either a single-component fit function
\begin{equation*}
\theta_K (T) = \theta_0 (1 - ((T+\Delta T) / T_C)^2) + c
\end{equation*}
or a two-component fit function
	\begin{align*}
&\theta_K (T) = \\
&\theta_0 \sqrt{[1 - ((T+\Delta T) / T_{C1})^2][1 - ((T+\Delta T) / T_{C2})^2]} + c
\end{align*}
When available, multiple data sets at the same optical power were fit simultaneously using the procedure available in \cite{multifit}. These fits were constrained to have identical values of $\Delta T$ and $\theta_0$ but were allowed to have independent values of the offset $c$, as this may vary over days due to small offset voltages in the measurement electronics. Error bars in plots of $\Delta T$ (and thus $T_K$) are the 95\% confidence intervals given by the fit procedures.

In order to calculate $T_K(0)$, a linear fit was performed on the extracted $T_K(P)$. A specialized fit procedure was used to correctly weight the data points according to their confidence intervals and to propagate the error \cite{York2004}. The values of $T_K(0)$ cited are those calculated from these fits, with error bars showing the 95\% confidence intervals calculated using this error propagation.

\section{Sample Survey}
\begin{figure}
	\includegraphics[]{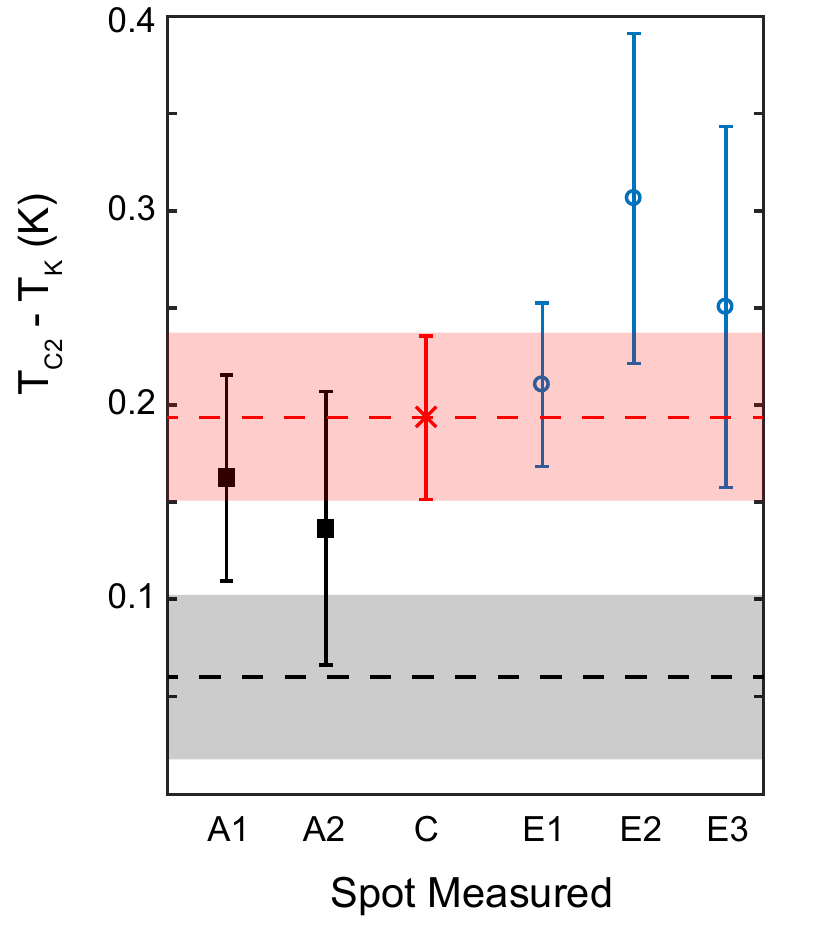}
	\caption{\label{deltats} Difference between lower superconducting transition temperature measured by AC susceptibility and Kerr temperature extracted from fits, each under 20 $\mu$W illumination. Horizontal axis labels indicate sample (letter) and spot (number) measured. Error bars are 95\% confidence intervals on $T_K$ extracted from fits. The red dashed line is the value expected if $T_K(0) = T_{C2}$, assuming that the optical heating is the same as measured on sample C; the red shaded region is the 95\% confidence interval on this value. The black dashed line is the value expected if instead $T_K(0) = T_{C1}$, and the grey shaded region is the 95\% confidence interval on this value. The data show that all spots except A2 are significantly outside the grey region, and A2 still overlaps much more with the red region, indicating that $T_K(0) = T_{C2}$. Note that measurement E2, which is an outlier \emph{above} the red region, was performed with exceptionally good optical focus, which may have increased the heating.}
\end{figure}

Three samples were measured in detail at 20 $\mu$W optical power: samples A, C, and E. Sample A was measured on 3 spots, sample C on 2, and sample E on 3. One measurement of sample A was corrupted by a damaged RF cable (it was undamaged for prior measurements and fixed for subsequent runs), and one measurement of sample C was corrupted by electrical noise from a nearby lab. In these cases the measured SNR was drastically lowered and thus the data was unusable; the other 6 measurements were reliable with sufficient SNR to fit for $T_K$. We label the measurements A1, A2, C, E1, E2, and E3. The samples were different sizes and so we expect different bulk heating due to the optical beam; indeed, we measured much lower AC susceptibility transition temperatures on sample C ($T_{C2}(20 \mu\mathrm{W}) = 1.433 K$) than on the larger samples A ($T_{C2}(20 \mu\mathrm{W}) = 1.568 K$) and E ($T_{C2}(20 \mu\mathrm{W}) = 1.477 K$). However, we expect that the difference between the temperature of the ``hot spot" under the optical beam and this bulk temperature will be the same across all samples. Finite element modeling of the hot spot shows that it only extends a few tens of microns and is thus unaffected by sample size; optical focus was similar for all measurements (except E2, discussed below) and so the hot spot temperature difference should be constant. We thus plot the difference between the $T_{C2}$ measured from AC susceptibility and the $T_K$ extracted from our fitting procedures. Data for these runs are shown in Fig.~\ref{deltats}.

Because we have detailed power dependence indicating $T_K(0) = T_{C2}$ for sample C, we may use the value of $T_{C2}(20 \mu\mathrm{W}) - T_K(20 \mu\mathrm{W})$ as a reference. If the hot spot temperature difference is the same for all measurements and the relevant transition temperature is always the lower transition, then we expect $T_{C2}(20 \mu\mathrm{W})-T_K(20 \mu\mathrm{W}) = 190 \pm 42$ mK. On the other hand if the relevant transition temperature is the upper transition, then $T_{C2}(20 \mu\mathrm{W}) - T_K(20 \mu\mathrm{W}) = 60 \pm 42$ mK, indicated by the dashed line and shaded region in the figure. We see that all measurements except A2 fall outside this region, with A2 still outside at the 2-$\sigma$ level when the error bars are combined. Note that measurement E2 appears significantly above the rest of the data points; this measurement was performed with higher SNR, indicating a tighter optical spot and likely a greater optical heating. The SNR was only 15\% higher, however, and the hot spot temperature difference is expected to scale no faster than SNR$^2$, so it is unlikely that this measurement corresponds to $T_K(0) = T_{C1}$. We thus take these measurements as moderate evidence that $T_K(0) = T_{C2}$ everywhere on the surface of these samples.

\begin{figure}
	\includegraphics[]{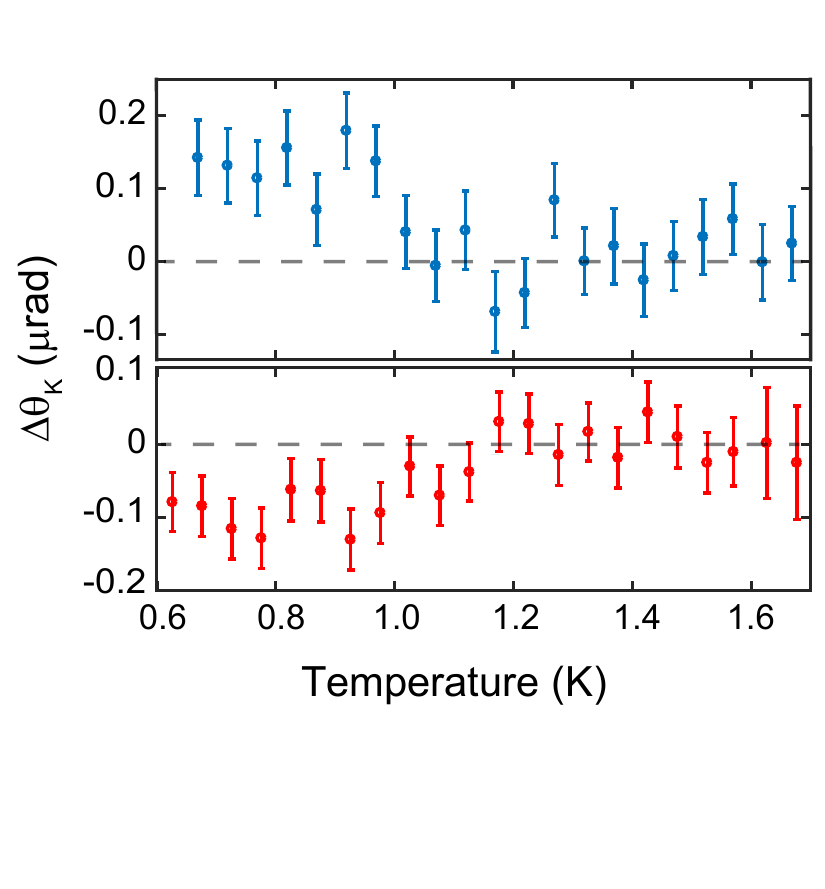}
	\caption{\label{samplesbd} Kerr angle as a function of temperature for Samples B (top, blue) and D (bottom, red)at 38 $\mu$W optical power. Sample B was trained in a +50 G field, while Sample D was trained in a -150 G field. Both show clear TRSB below $\sim$1.1 K, but the Kerr angle quickly saturates due to sample heating effects under this high-power illumination.}
\end{figure}

Samples B and D were measured at an earlier time, before we had developed our techniques for carefully accounting for optical heating. Both samples showed TRSB, as shown in Fig.~\ref{samplesbd}. However, data was not taken carefully at 20 $\mu$W, and so they cannot be used for our analysis of $T_K$. Note that Sample B was grown in a different batch, but has similar properties to the others.

\bibliographystyle{apsrev4-1}
\bibliography{suppbib}